\documentclass[a4paper]{spie}  %>>> use for US letter paper
\usepackage{graphicx}
\usepackage{amssymb, amsmath, subfigure}
\DeclareMathOperator{\Cum}{Cum}
\DeclareMathOperator{\E}{E}
\providecommand{\norm}[1]{\lvert#1\rvert}
\title{Techniques for noise removal from EEG, EOG and air flow signals in sleep 
patients} 
\author{Matthew J. Berryman\supit{a}, Sheila Messer\supit{a,b}, Andrew Allison\supit{a}, and Derek Abbott\supit{a}
\skiplinehalf
\supit{a}Centre for Biomedical Engineering and\\
School of Electrical and Electronic Engineering,\\
The University of Adelaide, SA  5005, Australia.\\
\supit{b}Texas Instruments Incorporated,\\
141 Stony Circle, Suite 130, MS 4110\\
Santa Rosa, California 95401, USA}
\authorinfo{Send correspondence to Derek Abbott\\
E-mail: dabbott@eleceng.adelaide.edu.au, Telephone: +61 8 8303 5748}
\begin{document} 
\maketitle
\begin{abstract}
Noise is present in the wide variety of signals obtained from sleep patients. This noise comes from a number of sources, from presence of extraneous signals to adjustments in signal amplification and shot noise in the circuits used for data collection. The noise needs to be removed in order to maximize the information gained about the patient using both manual and automatic analysis of the signals. Here we evaluate a number of new techniques for removal of that noise, and the associated problem of separating the original signal sources.
\end{abstract}
\section{INTRODUCTION}
Electroencephalograph (EEG) and electrooculograph (EOG) measurement techniques provide valuable information on sleep disorders~\cite{breathinglearning,EEGpredictor}. Recent studies looking at memory and learning during sleep have used these techniques as predictors of waking performance~\cite{breathinglearning,PFCsleep,memcons}. Comparison of thoracic and abdominal movements associated with breathing can reveal important information about breathing disorders and events such as apneas and hypopneas during sleep~\cite{autoparadoxical,TAasync,QSbreathing}. 

The process by which the EEG and EOG signals are recorded is described by Telpan~\cite{measurement}, but a brief summary is given here. The EEG and EOG signals are recorded by placing electrodes on the patient's scalp. These detect electric potentials generated by the flow of ions in neural cells that set up electric dipoles between the body of the neuron (soma) and the neural branches (apical dendrites). For the data we used, these signals were amplified, then digitized at 250 Hz for the EEG and 50 Hz for the EOG signal. We estimate the mutual information between the second EEG channel, from the left anterior position E1 to just below the opposite ear, with the left EOG channel, from just to the side of the left eye to the position just above the nose between the eyes. The signals are broken down into (typically) 30 second long epochs, these are then classified by a human operator into various stages of sleep and wakefulness. 

The main problems with analyzing the EEG and EOG signal are:
\begin{itemize}
\item Notch filtering of the the 50 Hz interference ripple from the signals also removes useful information.
\item Skin conductances can vary over time in different ways in different locations (however the gel used helps prevent this problem).
\item Due to conductances across the skin, the signal received by an electrode is a mixture of the true signals one is trying to measure.
\end{itemize}
The most significant problem is the mixing of signals, this can be reduced using blind signal separation techniques using higher order statistics~\cite{Belouchrani1,Belouchrani2}. The noise can then be removed using wavelet transforms~\cite{Matalg94,Bertra94,Lim95}.
We also consider these techniques for the thoracic and abdominal movements, for which similar problems may arise~\cite{QSbreathing}. 

\begin{figure}[htbp]
\centering
\subfigure[The time and power spectrum plots for the first eight seconds of the EEG data. Note there are many higher frequency signals superimposed on lower frequency signals, giving the appearance of noise, however this is important signal information that needs to be preserved. Note that some of the EOG signal is present on this signal.]{
  \includegraphics[width=7cm]{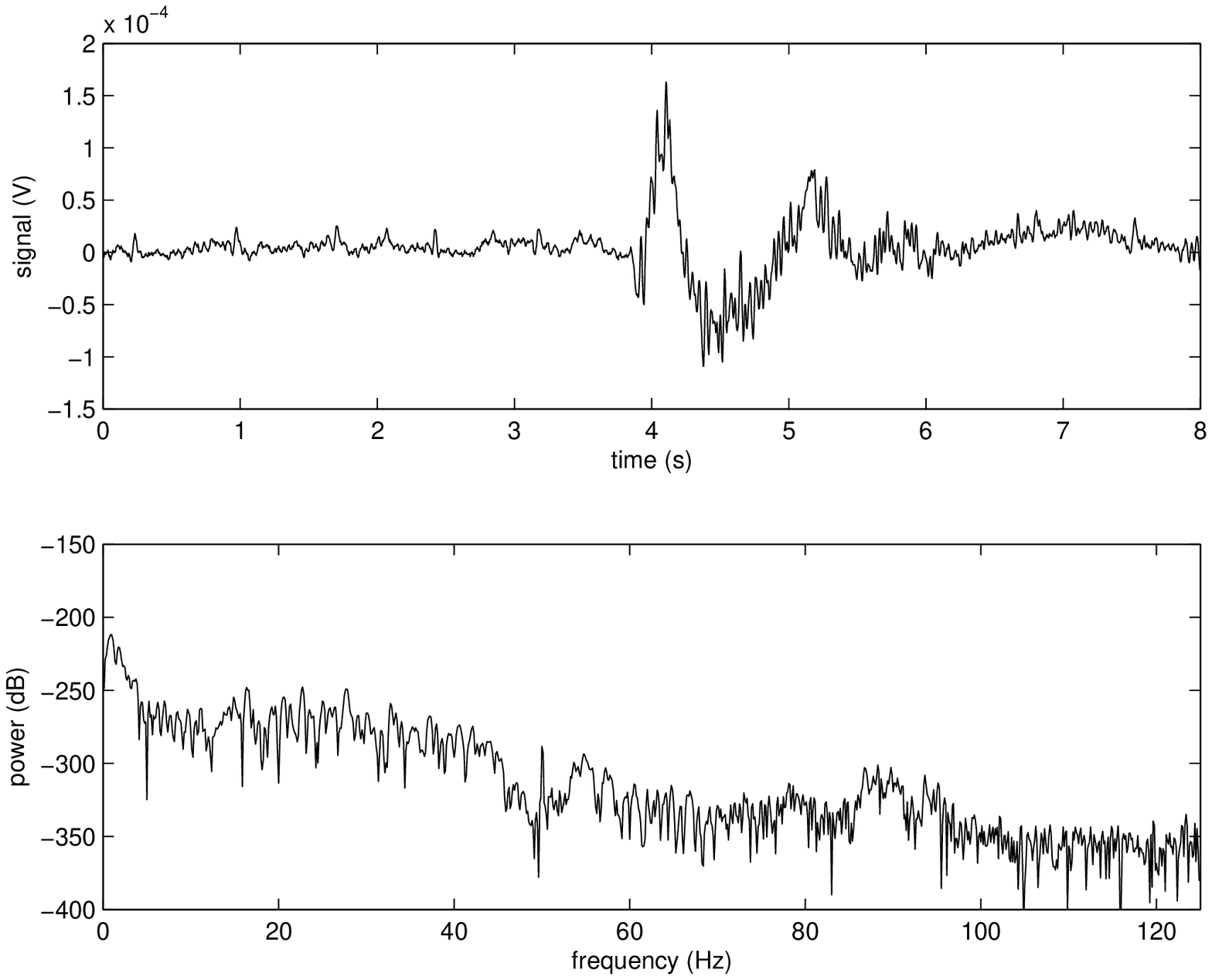}
  \label{timepowerEE:a}
  }
    \subfigure[The time and power spectrum plots for the first eight seconds of the EOG data. Note the lack of high frequency signals present in the EEG signal, since we are concerned here with low frequency muscle movement signals. The sampling rate used was correspondingly lower (50 Hz as opposed to 250 Hz for the EEG).]{
  \includegraphics[width=7cm]{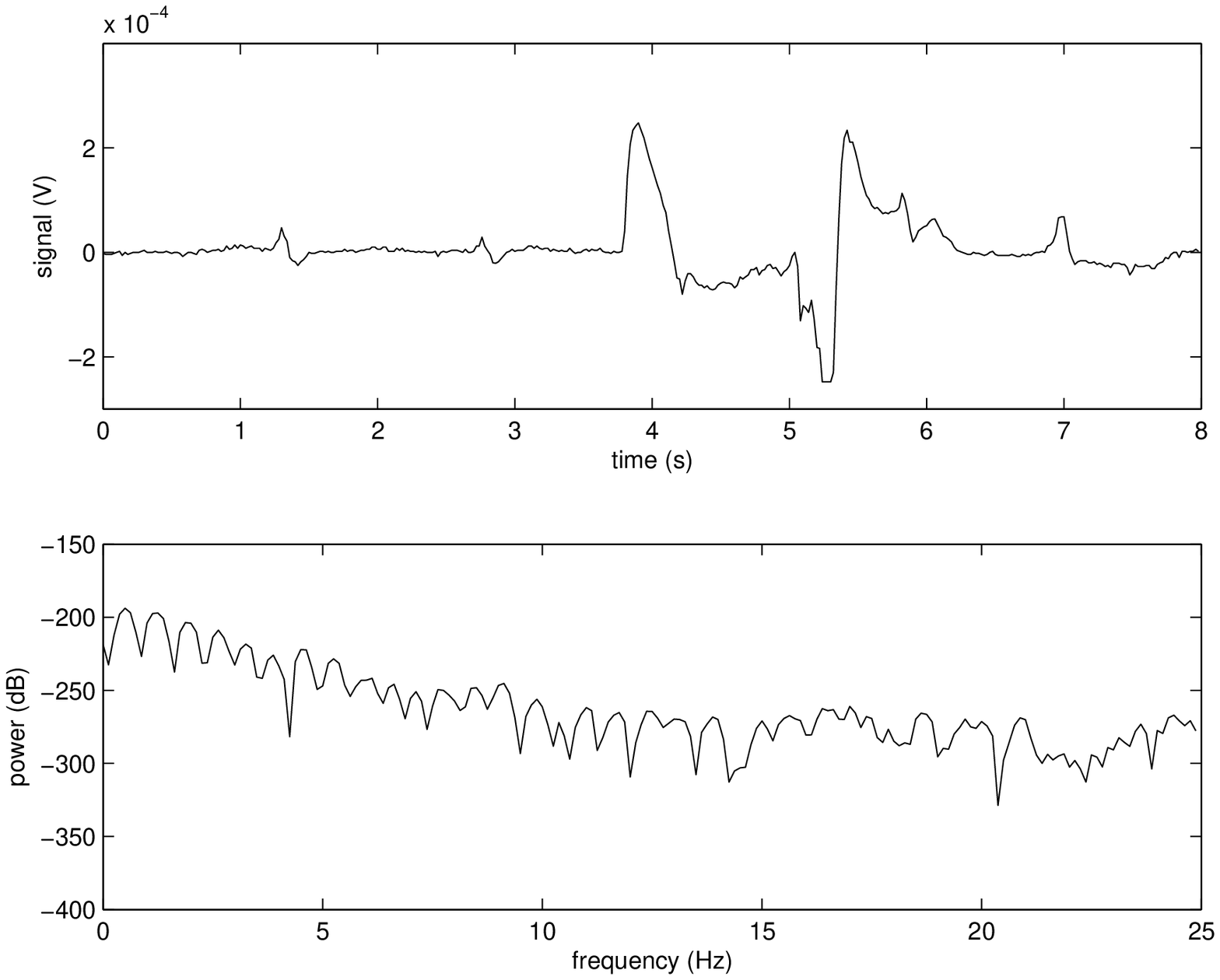}
  \label{timepowerEE:b}
  }
  \caption{The time and power spectra plots for the first eight seconds of the EEG and EOG data. Note the spectral differences between the two, with the EEG having many higher frequency components.}
\label{timepowerEE}
\end{figure}

\begin{figure}[htbp]
\centering
\subfigure[Time and power spectrum plots for the first eight seconds of the thoracic breathing data.]{
  \includegraphics[width=7cm]{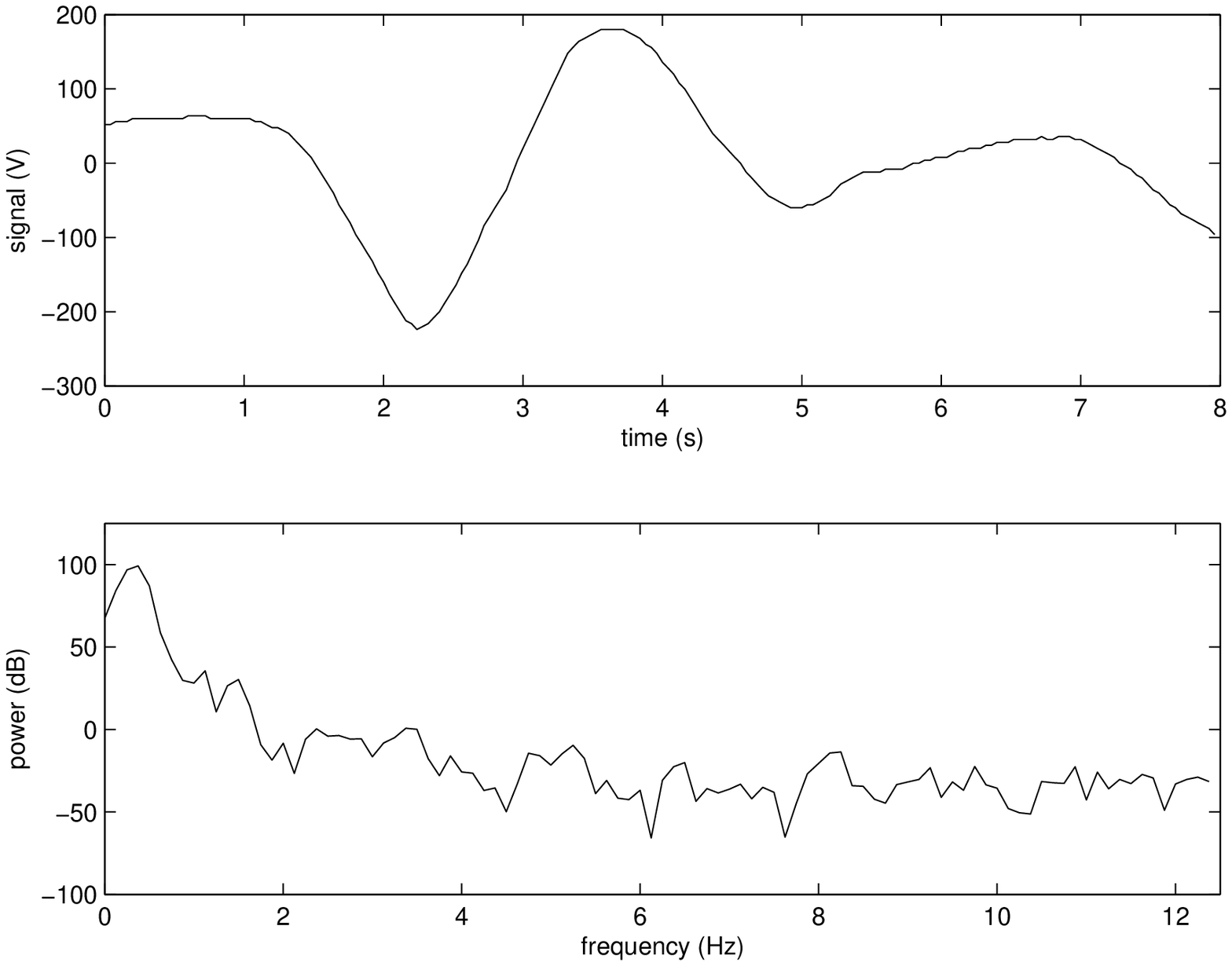}
  \label{timepowerTA:a}
  }
    \subfigure[Time and power spectrum plots for the first eight seconds of the abdominal breathing data.]{
  \includegraphics[width=7cm]{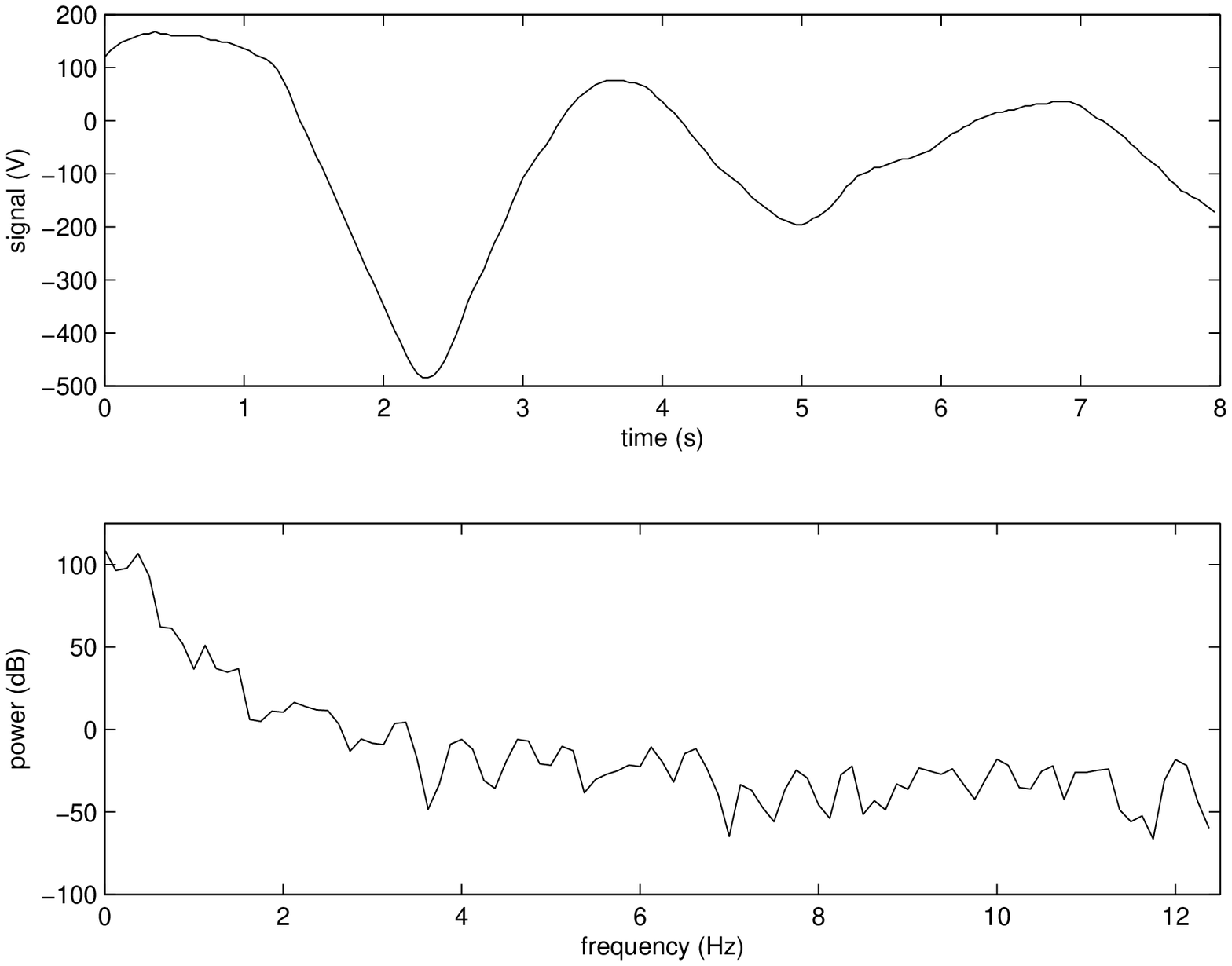}
  \label{timepowerTA:b}
  }
  \caption{Time and power spectra plots for the first eight seconds of the thoracic and abdominal breathing data. The sampling rate was 25 Hz, allowing the capture of relatively slow breathing signals. Note the thoracic signal has a slight phase lead over the abdominal breathing signal.}
\label{timepowerTA}
\end{figure}

Time and power spectra plots for one of the EOG and EEG sets of data is shown in Figure~\ref{timepowerEE}. Those for one of the thoracic and abdominal sets of data are shown in Figure~\ref{timepowerTA}.
\section{METHODS}
\begin{figure}[htbp]
  \centering{\resizebox{6cm}{!}{\includegraphics{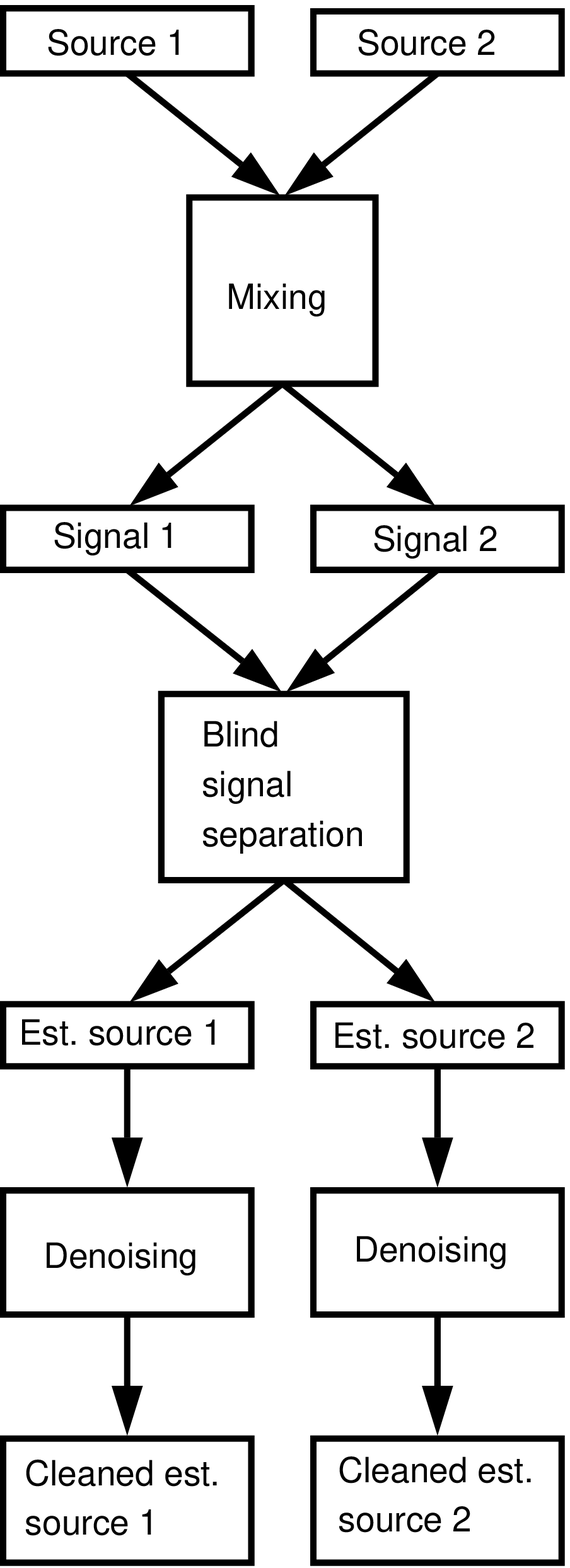}}}
\caption{Flowchart showing the general steps involved in going from the original sources to the cleaned, estimated sources. The optional Gaussian smoothing step is not shown.}
\label{flowchart}
\end{figure}

There are several problems to solve in eliminating noise from the signals. For the problem of the EOG and EEG signals, we must first make ensure the EOG and EEG signals have the same number of data points. To do this we use a Gaussian smoothing procedure, that is detailed in subsection~\ref{Gauss}. Other related smoothing procedures could be used, here we assume the distribution of the signal data is Gaussian, this is reasonably true for the data we use. We then have the problem of separating the sources from the observed signals, which contain a mixture of both. Three algorithms for this, detailed in subsections~\ref{SOBI} to~\ref{JCC}. The noise is removed from both the sources using wavelet transforms as elaborated on in subsection~\ref{waveletss}. A flowchart of the process is shown in Figure~\ref{flowchart}.

To evaluate the performance of the blind signal separation used to separate the source data, and to evaluate the noise removal we use an efficient algorithm, given in subsection~\ref{MIsubsec} to estimate the mutual information between two signals. We expect this measure to decrease when comparing the original signals with the separated signals, assuming greater independence between the separated signals, and to remain the same when comparing the noisy signals with those where the noise has been removed, assuming the noise is uncorrelated between the two signals. This is largely true for the signals of interest, although there may be some information at certain frequencies that is correlated due to extraneous electromagnetic signals being received by the leads, as they act as antennas. This is kept to a minimum through appropriate grounding. 
\subsection{Gaussian smoothing}
\label{Gauss}
The EEG signal has a sampling rate five times higher than the EOG (250 Hz to 50 Hz). These are recorded simultaneously, so every fifth time point in the EEG corresponds to a time point of the EOG signal. We use Gaussian smoothing to reduce the number of data points in the EEG by a factor of five:
\begin{equation}
s\left(i\right)=\displaystyle\frac{\displaystyle\sum_{j=i-m}^{i+m}x_{j}w\left(x_{j},x_{i}\right)}{\displaystyle\sum_{j=i-m}^{i+m}w\left(x_{j},x_{i}\right)},
\label{gsmooth}
\end{equation}
where the weights $w\left(x_{j},x_{i}\right)$ are
\begin{equation}
w\left(x_{j},x_{i}\right)=e^{-\left(x_{i}-x_{j}\right)^{2}/\left(2\hat{\sigma}^{2}\right)},
\label{weight}
\end{equation}
$i$ is the discrete time point we are calculating the smoothed average for, and $\hat{\sigma}^{2}$ is the estimate of variance for the entire set of samples in the EEG signal. Now that the two sets of data, which can be written as $\mathbf{x}\left(t\right)=\left[x_{1}\left(t\right),x_{2}\left(t\right)\right]^{T}$, we can apply blind signal separation, using the following model of the data.
\subsection{Data model for blind signal separation}
We write the original, $m$-dimensional source data as $\mathbf{s}\left(t\right)=\left[s_{1}\left(t\right),s_{2}\left(t,\right),\ldots,s_{m}\left(t\right)\right]^{T}$. It is assumed that the sources are independent. We then consider an unknown linear model $\mathbf{A}_{n\times m}$ generating the observed signals, written as an $n$-dimensional vector $\mathbf{x}\left(t\right)=\left[x_{1}\left(t\right),x_{2}\left(t\right)\ldots,x_{n}\left(t\right)\right]^{T}$
by
\begin{equation}
\mathbf{x}\left(t\right) = \mathbf{As}\left(t\right),
\label{linearmodel}
\end{equation}
where $\mathbf{A}$ is referred to as the mixing matrix. Being able to swap columns of $\mathbf{A}$, and scaling a source by a scaling change in a row of $\mathbf{A}$ means there is an ambiguity in both the permutation (of labeling) of the sources and the scaling of the sources respectively. With this model of the data, we can now apply blind signal separation techniques.
\subsection{Blind signal separation}
There are two key blind signal separation approaches that are combined to form the joint cumulant and correlation (JCC) algorithm in subsection~\ref{JCC}. They are the second order blind identification (SOBI) algorithm, discussed in subsection~\ref{SOBI}, and the joint approximate decomposition of eigenmatrices (JADE), discussed in subsection~\ref{JADE}. Both approaches have a common first step, in which the data is whitened using a {\it sphering} matrix $\mathbf{W}$, which transforms the mixing matrix $\mathbf{A}$ into a unitary matrix $\mathbf{U}$, which is a matrix for which $\mathbf{U}\mathbf{U}^{T}=\mathbf{I}$~\cite{Belouchrani1,Belouchrani2}. The next step of estimating $\mathbf{A}$ is dependent on the choice of algorithm and detailed below.  
\subsection{SOBI algorithm}
\label{SOBI}
Given a hypothesis of sources with different spectra and the linear model of Equation~\ref{linearmodel}, we can calculate time-delayed, cross-correlation matrices,
\begin{equation}
\begin{split}
\mathbf{R}\left(\tau\right) &=\E\left[\mathbf{x}\left(t\right)\mathbf{x}\left(t-\tau\right)^{T}\right]\\
& \mathbf{A}\mathbf{R}_{s}\left(\tau\right)\mathbf{A}^{H}.
\end{split}
\label{R}
\end{equation}
for $\tau\neq 0$ and 
\begin{equation}
\mathbf{R}_{s}=
\begin{pmatrix}
\E\left[s_{1}\left(t\right)s_{1}\left(t-\tau\right)\right] & 0 & 0 & \cdots & 0\\
0 & \E\left[s_{2}\left(t\right)s_{2}\left(t-\tau\right)\right] & 0 & \cdots & 0\\
\vdots & 0 & \ddots & & \vdots\\
0 & \ldots & & 0 & \E\left[s_{m}\left(t\right)\left(t-\tau\right)\right]
\end{pmatrix},
\label{Rs}
\end{equation}
for $E[\cdot]$ the expectation operator. The correlation matrices can then be whitened,
\begin{equation}
\underline{\mathbf{R}}=\mathbf{W}\mathbf{R}\left(\tau\right)=\mathbf{U}\mathbf{R}_{s}\left(\tau\right)\mathbf{U}^{T},
\label{whitecorr}
\end{equation}
$\forall t \neq 0$.
The joint diagonalization of the set of $p$ whitened correlation matrices $\left\{\underline{\mathbf{R}}\left(\tau_{i}\right)\lvert i=1,\ldots,p\right\}$~\cite{Belouchrani1}. The matrix $\mathbf{U}$ can only be uniquely determined iff for any $(i,j)$, there exists at least one lag $\tau_{k}$ such that $\E\left[s_{i}\left(t\right)s_{i}\left(t-\tau\right)\right] \neq \E\left[s_{j}\left(t\right)s_{j}\left(t-\tau\right)\right]$~\cite{Belouchrani1}. The mixing matrix is then estimated by $\hat{\mathbf{A}}=\mathbf{W}\mathbf{U}$. An alternative to the SOBI algorithm is the JADE algorithm.
\subsection{JADE algorithm}
\label{JADE}
Here we assume the linear model of Equation~\ref{linearmodel} and assume independence of sources.
To each $n$-dimensional vector $x$ is associated a quadicovariance matrix $\mathbf{Q}:\mathbf{M}\rightarrow\mathbf{N}$ defined by $\mathbf{N}=\mathbf{QM}$ such that
\begin{equation}
N_{i,j}=\displaystyle \sum_{\left(k,l\right)} \Cum\left(x_{i},x_{j},x_{k},x_{l}\right) M_{k,l},
\label{Q}
\end{equation}
where $\Cum\left(\cdot\right)$ is defined as 
\begin{equation}
\Cum\left(x_{i},x_{j},x_{k},x_{l}\right)=\E\left[\bar{x}_{i}\bar{x}_{j}\bar{x}_{k}\right] -
\E\left[\bar{x}_{i}\bar{x}_{j}\right]\E\left[\bar{x}_{k}\bar{x}_{l}\right]-
\E\left[\bar{x}_{i}\bar{x}_{k}\right]\E\left[\bar{x}_{j}\bar{x}_{l}\right]-
\E\left[\bar{x}_{i}\bar{x}_{l}\right]\E\left[\bar{x}_{j}\bar{x}_{k}\right],
\label{Cum}
\end{equation}
and where $\bar{x}_{i}=x_{i}-\E\left[x_{i}\right]$, etcetera~\cite{Cardoso2}. As the set of $n\times n$ matrices is an $n^{2}$-dimensional linear space, it can be shown that there exist $n^{2}$ real eigenvalues $\lambda_{r}$ and $n^{2}$ orthonormal eigenmatrices $\mathbf{M}_{r}$ satisfying $\mathbf{QM}_{r}=\lambda_{r}\mathbf{M}_{r}$~\cite{Belouchrani1}. It can be proved that only $n$ of the eigenvalues are non-zero~\cite{Cardoso1}, and joint diagonalizaton of the $n$ corresponding eigenmatrices, labeled $\underbar{M}_{r}$, gives the unitary matrix $U$~\cite{Belouchrani1}. As with the SOBI algorithm, the mixing matrix is estimated by $\hat{\mathbf{A}}=\mathbf{WU}$. Combining the two algorithms gives us the JCC algorithm.
\subsection{JCC algorithm}
\label{JCC}
In JCC, we use both the correlation information provided by the SOBI algorithm of subsection~\ref{SOBI},  $\underline{\mathbf{R}}\left(\tau_{i}\right)$, and the cumulant quadricovariance eigenmatrices, $\underline{\mathbf{M}}_{r}$, provided by the JADE algorithm. Joint diagonalization gives the unitary matrix $\mathbf{U}$, which again acts to give an estimator $\underline{\mathbf{A}}=\mathbf{WU}$. Using $\hat{\mathbf{A}}$ we separate the signals into estimates of the original source data. We then consider removing the noise from this data, using wavelet techniques. 
\subsection{Wavelet noise removal}
\label{waveletss}
The mathematical description of the continuous wavelet transform (CWT) of $f\in\mathbf{L^{2}}\left(\mathbb{R}\right)$
is described by Mallat~\cite{Mallat99} as 
\begin{equation}
\left(Wf\right)\left(u,s\right)=\int_{-\infty}^{+\infty}f\left(t\right)\psi_{u,s}^{*}\left(t\right)dt,
\label{eqn:cwt}
\end{equation}
where 
\begin{equation}
\psi_{u,s}\left(t\right)=\frac{1}{\sqrt{s}},
\label{eqn:wavelet}
\end{equation}
is a family of orthogonal wavelets, $\norm{\psi_{u,s}} = 1$, $\langle \psi_{u,s} ,\psi_{u',s'}\rangle = 0$ for $\left(u,s\right)\neq \left(u',s'\right)$, and
\begin{equation}
\int_{-\infty}^{+\infty}\psi_{u,s}\left(t\right)dt=0.
\label{eqn:zeroavg}
\end{equation}
The scale of the wavelet may conceptually be considered the
inverse of the frequency.  

\begin{figure}[htbp] 
\centering
\includegraphics[width=7cm]{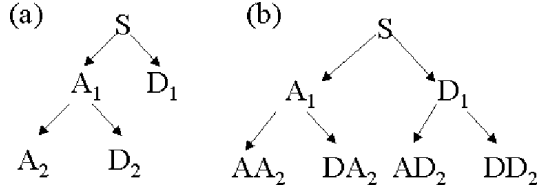}
\caption{This figure illustrates how (a) the discrete
  wavelet transform decomposes a signal into details and
  approximations iteratively decomposing the approximations; 
  (b) wavelet packets iteratively decompose the approximations and details.}
\label{figure2} 
\end{figure}

The CWT reveals much detail about a signal, however due to the continuous nature it cannot be computed for real signals on a digital computer. 
Therefore, the discrete wavelet transform (DWT) is normally used.
The DWT calculates the wavelet coefficients at discrete intervals of time and
scale instead of at all scales. With the DWT, a fast version of the algorithm is
possible, analogous to the fast Fourier transform. This version of the algorithm makes use of the fact that if scales and positions are chosen based on powers of two (dyadic scales and positions) the analysis is
 very efficient. In 1988, Mallat developed an efficient way to implement this algorithm, which is
known as a two-channel sub-band coder~\cite{Mallat89}. 
For a single level of decomposition, this algorithm passes the signal through
two complementary (high-pass and low-pass) filters resulting in
approximations which are high-scale, low-frequency components of the
signal, and details, which are low-scale, high-frequency components of
the signal. This results in twice as many data-points so the data is
down-sampled.  For further levels of decomposition, successive
approximations may be iteratively broken down into details and approximations
as shown in Figure~\ref{figure2}. Coefficients below a certain level are regarded as noise and
thresholded out.  Thresholding may be soft or hard.  Hard thresholding
 is defined as
\begin{equation}
\begin{array}{cccc}
y & = & x & \qquad \textrm{for}\:|x|>\theta \\
y & = & 0 & \qquad \textrm{for}\:|x|\leq \theta
\end{array}
\label{eqn:hard_thresholding}
\end{equation}
and soft thresholding as
\begin{equation}
\begin{array}{cccc}
y & = &\textrm{sign}(x)(|x|-\theta) &\qquad \textrm{for}\:|x|>\theta \\
y& =& 0 &\qquad \textrm{for}\:|x|\leq \theta
\end{array}
\label{eqn:soft_thresholding}
\end{equation}
where $x$ is the original signal, $y$ is the thresholded signal, and
$\theta$ is the threshold. Hard thresholding tends to create
discontinuities at $x=\pm \theta$ because any values of the signal less than
the threshold are immediately set to zero. With soft thresholding, the
thresholded values are shrunk towards zero without creating the
discontinuities.  The signal is then reconstructed without significant
 loss of information. Then the signal may be reconstructed by up-sampling,
passing the approximations and details through the appropriate reconstruction
filters and combining the results. Based on SNR measures of wavelet performance, we used Daubechies wavelets of order 5, with soft thresholding and a decomposition level of 5; although this is not the best for noise removal, we are more interested in preservation of information when going from the estimated sources to the denoised estimated sources.

To evaluate the performance of the above techniques we introduce a measure of mutual information.
\subsection{MI estimation algorithm}
\label{MIsubsec}
The following is an outline of the method we use in calculating the mutual information between the EEG and EOG signals, as given in Kraskov {\it et al.}~\cite{Grassberger}. 

Mutual information for two signals $X$ and $Y$ is defined in  Equation~\ref{MI}
\begin{equation}
I\left(X,Y\right) = \displaystyle\int_{-\infty}^{\infty}\int_{-\infty}^{\infty}\mu\left(x,y\right)\log\frac{\mu\left(x,y\right)}{\mu_{x}\left(x\right)\mu_{y}\left(y\right)}dxdy,
\label{MI}
\end{equation}
where $\mu$, $\mu_{x}$ and $\mu_{y}$ are probability measures. 
We then take the set of points $z_{i}=\left(x_{i},y_{i}\right)$ for the EEG $x_{i}$ and EOG $y_{i}$, $i=1,\ldots\,N$ . Then we find the $k$th closest neighbor of each $z_{i}$ according to the metric 
\begin{equation}
\norm{z-z'}=\max\{\norm{x-x'},\norm{y-y'}\}.
\label{metric}
\end{equation}
The $k$th nearest neighbor is then projected onto the $x$ and $y$ axes giving the distances $\epsilon_{x}\left(i\right)/2$ and $\epsilon_{y}\left(i\right)/2$ respectively. The mutual information is estimated by:
\begin{equation}
\hat{I}_{k}\left(X,Y\right)\approx\psi\left(k\right)-1/k-\langle\psi\left(n_{x}\right)+\psi\left(n_{y}\right)\rangle+\psi\left(N\right),
\label{estimate}
\end{equation}
where $\psi\left(\cdot\right)$ is the digamma function given by
\begin{equation}
\psi\left(z\right)=\frac{d}{dz}\ln \Gamma\left(z\right),
\label{digamma}
\end{equation}
and 
\begin{equation}
\langle\ldots\rangle=\frac{1}{N}\displaystyle\sum_{i=1}^{N}\E\left[\ldots\left(i\right)\right].
\label{tavg}
\end{equation}
\section{RESULTS}
\subsection{Blind signal separation}
We compared the blind signal separation for three algorithms, across four sets of data, two of thoracic and abdomen (TA) breathing data, and two of EEG and EOG (EE) data. The differences in mutual information between the signal data and the estimated source data are shown in Table~\ref{BSSMI}. The higher the mutual information, the better the algorithm is for separating the original sources, given the assumptions of that algorithm.
\begin{table}[htbp]
\centering
\caption{The difference in mutual information (in nats/sample) between the two estimated sources and the two signals, $\hat{I}_{k}\left(\text{est. sources}\right)-\hat{I}_{k}\left(\text{signals}\right)$, for the two sets of thoracic-abdominal (TA) data and the two sets of EEG and EOG (EE) data. Nats are units of information, when a natural logarithm is used. This is computed for all three (SOBI, JADE, and JCC) algorithms}
\begin{tabular}{|c||c|c|c|c|}\hline
 & {\bf TA1} & {\bf TA2} & {\bf EE1} & {\bf EE2}\\\hline\hline
{\bf SOBI} & 1.0319 & 0.4539 & 0.4522 & 0.6101 \\\hline
{\bf JADE} & $<0.0001$ & $<0.0001$ & $<0.0001$ & $<0.0001$\\\hline
{\bf JCC} & 0.3640 & 0.4292 & 0.4032 & 0.5342 \\\hline
\end{tabular}
\label{BSSMI}
\end{table}
\subsection{Wavelet denoising}
For each of the generated estimates of the sources (three blind signal separation algorithms applied to four pairs of signals) we estimated the mutual information between the estimates of the sources, and the denoised estimates of the sources. These are given in Table~\ref{WDMI}. We observe no difference in mutual information between 
\begin{table}[htbp]
\centering
\caption{The difference in mutual information (in nats/sample) between the JADE estimated sources and the wavelet denoised estimated sources is computed for the two sets of thoracic-abdominal (TA) data and the two sets of EEG and EOG (EE) data, $\hat{I}_{k}\left(\text{est. denoised sources}\right)-\hat{I}_{k}\left(\text{est. sources}\right)$. Nats are units of information, when a natural logarithm is used.}
\begin{tabular}{|c|c|c|c|}\hline
{\bf TA1} & TA2 & EE1 & EE2\\\hline
$<0.0001$ & $<0.0001$& $<0.0001$ & $0.0001$\\\hline\hline
\end{tabular}
\label{WDMI}
\end{table}
\section{DISCUSSION \& CONCLUSIONS}
For our particular set of thoracic and abdominal breathing data, the SOBI algorithm works well, with an increase in the mutual information, probably because the sources have reasonably distinct spectra. Since the JCC combines information from both the SOBI and JADE algorithms by way of joint diagonalization, it introduces the problems associated with using the JADE algorithm for this data, namely that the sources are not independent. The two sources have a high level of dependence, being almost synchronous during regular breathing, tending to differ only for compliant chests in young children or when a breathing obstruction occurs~\cite{autoparadoxical,QSbreathing}.
Similarly, for the EEG and EOG data, although these are more independent, the SOBI algorithm performs best at separating the original sources from the observed signals.

The wavelet denoising performs well, in that it preserves (as far as we can determine) the information present in the signals. Further work will consider wavelet packet and matching pursuit denoising algorithms~\cite{Mallat89,Mallat93,Krishn00}, and how these effect mutual information between two different channels. We will also consider the effect of swapping the denoising and blind signal separation techniques, in theory this should have little to no difference on the results.
\section{ACKNOWLEDGEMENTS}
We acknowledge funding from The University of Adelaide. Useful discussions with Cosma Shalizi from the Center for the Study of Complex Systems at the University of Michigan were greatly appreciated.
\bibliography{phd}   %>>>> bibliography data in report.bib
\bibliographystyle{spiebib}   %>>>> makes bibtex use spiebib.bst
\end{document}